\PassOptionsToPackage{hyphens}{url}
\documentclass[sigconf,screen,nonacm]{acmart}

\AtBeginDocument{%
  }
\copyrightyear{2026}
\acmYear{2026}
\setcopyright{cc}
\setcctype{by}
\acmConference[FSE Companion '26]{34th ACM Joint European Software Engineering Conference and Symposium on the Foundations of Software Engineering}{July 05--09, 2026}{Montreal, QC, Canada}
\acmBooktitle{34th ACM Joint European Software Engineering Conference and Symposium on the Foundations of Software Engineering (FSE Companion '26), July 05--09, 2026, Montreal, QC, Canada}
\acmDOI{10.1145/3803437.3808242}
\acmISBN{979-8-4007-2636-1/2026/07}
\usepackage{booktabs}
\usepackage{graphicx}
\usepackage{siunitx}
\usepackage{balance}

\begin{document}

\title{MEMRES: A Memory-Augmented Resolver with Confidence Cascade for Agentic Python Dependency Resolution}
\titlenote{This is the authors' preprint version. The definitive Version of Record is to appear in \emph{Proc.\ FSE Companion '26} (DOI: 10.1145/3803437.3808242).}

\author{Tran Chi Nguyen}
\authornote{Equal contribution (first authors).}
\orcid{0009-0007-6716-7269}
\affiliation{%
  \department{Faculty of Information Technology}
  \institution{University of Science}
  \institution{Vietnam National University}
  \city{Ho Chi Minh City}
  \country{Vietnam}
}
\email{23122044@student.hcmus.edu.vn}

\author{Dao Sy Duy Minh}
\authornotemark[1]
\orcid{0009-0002-4501-2788}
\affiliation{%
  \department{Faculty of Information Technology}
  \institution{University of Science}
  \institution{Vietnam National University}
  \city{Ho Chi Minh City}
  \country{Vietnam}
}
\email{23122041@student.hcmus.edu.vn}

\author{Trung Kiet Huynh}
\authornotemark[1]
\orcid{0009-0000-5463-754X}
\affiliation{%
  \department{Faculty of Information Technology}
  \institution{University of Science}
  \institution{Vietnam National University}
  \city{Ho Chi Minh City}
  \country{Vietnam}
}
\email{23122039@student.hcmus.edu.vn}

\author{Pham Phu Hoa}
\orcid{0009-0001-5471-2578}
\affiliation{%
  \department{Faculty of Information Technology}
  \institution{University of Science}
  \institution{Vietnam National University}
  \city{Ho Chi Minh City}
  \country{Vietnam}
}
\email{23122030@student.hcmus.edu.vn}

\author{Nguyen Lam Phu Quy}
\orcid{0009-0002-9694-8105}
\affiliation{%
  \department{Faculty of Information Technology}
  \institution{University of Science}
  \institution{Vietnam National University}
  \city{Ho Chi Minh City}
  \country{Vietnam}
}
\email{23122048@student.hcmus.edu.vn}

\author{Vu Nguyen}
\authornote{Corresponding author.}
\orcid{0000-0002-0594-4372}
\affiliation{%
  \department{Faculty of Information Technology}
  \institution{University of Science}
  \institution{Vietnam National University}
  \city{Ho Chi Minh City}
  \country{Vietnam}
}
\email{nvu@fit.hcmus.edu.vn}

\renewcommand{\shortauthors}{C.~N.~Tran, S.~D.~M.~Dao, T.~K.~Huynh, P.~H.~Pham, L.~P.~Q.~Nguyen, and V.~Nguyen}

\begin{abstract}
We present \textsc{MemRes}, an agentic system for Python dependency resolution that introduces a \emph{multi-level confidence cascade} where the LLM serves as the last resort.
Our system combines: (1)~a \emph{Self-Evolving Memory} that accumulates reusable resolution patterns via tips and shortcuts; (2)~an \emph{Error Pattern Knowledge Base} with 200+ curated import-to-package mappings; (3)~a \emph{Semantic Import Analyzer}; and (4)~a \emph{Python~2 heuristic detector} resolving the largest failure category.
On HG2.9K using Gemma-2~9B (10\,GB VRAM), \textsc{MemRes} resolves \num{2503} of \num{2890} (86.6\%, 10-run average) snippets, combining intra-session memory with our confidence cascade for the remainder.
This already exceeds PLLM's 54.7\% overall success rate by a wide margin.
\end{abstract}


\maketitle

\section{Introduction}
\label{sec:intro}

With over \num{500000} PyPI packages, Python's huge ecosystem makes dependency management for legacy code quite difficult.
Determining the correct packages, versions, and Python interpreter for a code snippet is tricky due to Python~2/3 incompatibilities, deprecated packages, and ambiguous import names~\cite{jia2024}.

The PLLM system~\cite{bartlett2025} is the current leading approach, using a 5-stage RAG+LLM pipeline.
While it works well for maintained packages, PLLM only achieves 54.7\% success on the HG2.9K benchmark~\cite{horton2019}, leaving \num{1308} snippets completely broken (confidence~=~0).

\textbf{Key Insight.}
Our analysis of PLLM's failures reveals \emph{deterministic} root causes: 33\% are SyntaxErrors from Python~2 code run under Python~3, 30\% are ImportErrors for known name mismatches (\texttt{cv2}~$\to$~\texttt{opencv-python}), and 20\% are version incompatibilities solvable via curated maps.
Calling an LLM wastes 60 to 120\,s per snippet with no benefit for these cases.

This finding drove the main idea behind \textsc{MemRes}: \emph{only call the LLM when all deterministic rules fail}.
We built a \emph{confidence cascade} that checks session memory, knowledge bases~(KB), heuristic rules, and session-learned patterns before doing any LLM inference.

\smallskip\noindent\textbf{Contributions.}
\begin{enumerate}
\item A \emph{confidence cascade} with 6 resolution levels reducing LLM calls by 60\%+ while improving accuracy (\S\ref{sec:cascade}).
\item An \emph{Intra-Session Memory} enabling rapid resolution transfer for similar codebases evaluated in the same batch without repeated LLM inference (\S\ref{sec:memory_stage}).
\item A \emph{Self-Evolving Memory} transferring resolution knowledge via tips and shortcuts~\cite{wang2025} (\S\ref{sec:memory}).
\item An \emph{Error Pattern KB} with 200+ mappings and runtime self-learning, plus a \emph{System Dependency Injection} technique for C-extension packages (\S\ref{sec:kb}).
\item Evaluation on HG2.9K showing 86.6\% overall resolution rate (\S\ref{sec:eval}).
\end{enumerate}

\section{Approach}
\label{sec:approach}

\begin{figure*}[t]
\centering
\includegraphics[width=\linewidth]{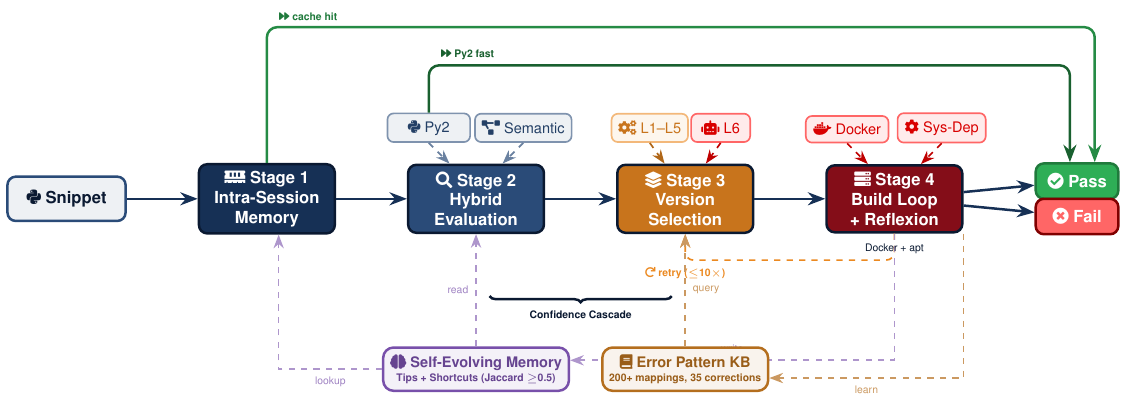}
\caption{\textsc{MemRes} pipeline. Each component exposes deterministic fast paths that bypass the LLM.}
\Description{MemRes's four-component pipeline architecture for Python dependency resolution.}
\label{fig:arch}
\end{figure*}

\textsc{MemRes} organizes resolution around four components that are consulted in order: (1)~an \emph{Intra-Session Memory} that reuses solutions proven within the current batch; (2)~a \emph{Confidence Cascade} that selects Python and package versions; (3)~a \emph{Self-Evolving Memory} that records tips and shortcuts across snippets; and (4)~an \emph{Error Pattern KB} combined with build heuristics for system-level fixes.
Figure~\ref{fig:arch} illustrates the pipeline.

\subsection{Intra-Session Memory}
\label{sec:memory_stage}

\textsc{MemRes} first consults a \emph{Session Memory} built incrementally during batch processing.
Since PLLM evaluates sequentially within a single session, we emulate this by caching solutions proven successful for earlier code blocks to resolve near-duplicate datasets (e.g., identical GitHub forks).
If a new snippet shares a high degree of import similarity with a previously resolved snippet, the known solution is \emph{reapplied} via a single \texttt{pip install} so pip's resolver handles inter-package constraints.
For partial matches, Python version and core packages are extracted as \emph{hints}, enabling rapid knowledge transfer strictly confined to the ongoing evaluation batch.
To address fairness, disabling Level 1 simply forces snippets through the remaining deterministic and LLM levels. Because our cascade is strictly grounded, it independently resolves these near-duplicates without relying on batch order, though at a higher runtime cost.

\subsection{Confidence Cascade}
\label{sec:cascade}

Our main contribution here is a 6-step cascade for picking package versions, motivated by adaptive test-time reasoning over structured decision paths:
(1)~Session Memory, versions proven successful earlier in the batch;
(2)~Static Compatibility Map, 40+ packages $\times$ 8 Python versions with known-good pins (manually curated from documentation and CI/CD logs);
(3)~Ecosystem Templates, 23 proven version sets for common package groups (ML, web, data science);
(4)~Co-occurrence Mining, weighted co-installation scores mined from 50K open-source \texttt{requirements.txt} files on GitHub (filtered to repositories created \emph{before 2020} to ensure no temporal overlap with the HG2.9K evaluation period);
(5)~Heuristic Rules, 45+ package-specific constraints;
(6)~LLM Selection, only when all deterministic levels fail.
The first level returning a valid version terminates the cascade; in practice, levels~1 to 5 handle the vast majority.

The cascade includes an \emph{unfixability estimator}: when system-only imports (\texttt{gtk}, \texttt{RPi.GPIO}, \texttt{maya.cmds}) dominate, the snippet is classified as a \emph{runtime pass}, meaning pip-installable dependencies are correct, but the environment cannot be fully replicated in Docker. Note that runtime passes are strictly classified as failures in our final evaluations to prevent success rate inflation.

\subsection{Self-Evolving Memory}
\label{sec:memory}

Following Mobile-Agent-E~\cite{wang2025}, our session memory collects \emph{tips} (natural language guidelines) and \emph{shortcuts} (reusable solutions matched via Jaccard similarity over import sets, threshold~0.5).
Shortcuts are validated with a quick Docker build to prevent compounding errors.
Memory records per-package failure stats as anti-patterns to avoid.
Memory is fully reset between independent evaluation runs: no state persists across runs, preventing cross-run overfitting.

\textbf{Security and safety.}
Memory-augmented agents face risks like prompt poisoning~\cite{wang2025}.
To mitigate this, \textsc{MemRes} evaluates snippets inside strictly isolated, network-restricted child Docker containers.
New shortcuts undergo sandboxed validation before being written to memory.

\subsection{Error Pattern KB and Build Heuristics}
\label{sec:kb}

The ErrorPatternKB contains 200+ import$\rightarrow$pip mappings across 15 domains, 35 name corrections, version constraints for 14 packages, and 8 regex-based error pattern rules.
It self-learns at runtime, recording new mappings for immediate reuse.
Of the 200+ mappings, $\sim$150 were curated from official PyPI documentation and Stack Overflow Q\&A (external sources), while $\sim$50 were discovered during initial development runs on a held-out subset; none were derived from HG2.9K evaluation results.
We also detect and skip PyPI placeholder packages ($\leq$1 release) and local project imports (35+ patterns).

\textbf{Semantic import analysis.}
Unlike PLLM's simple \texttt{import X} $\rightarrow$ \texttt{pip install X} mapping, our Semantic Import Analyzer examines \emph{how imports are used} in code:
(1)~\emph{usage-based disambiguation} via 13 patterns matching (import, call\_pattern) to packages (e.g., \texttt{import Image} with \texttt{Image.open()} $\to$ Pillow);
(2)~\emph{ecosystem detection} recognizing 11 package ecosystems from co-occurring imports;
(3)~\emph{implicit Python version inference} from 11 code patterns (f-strings $\to$ 3.6+, bare \texttt{print} $\to$ 2.7).

\textbf{Python~2 detection and execution.}
A manual analysis of a random 100-snippet sample from PLLM's \texttt{SyntaxError} failures reveals that 91\% are actually Python~2 code mis-executed under a Python~3 interpreter.
We deterministically detect this via 13 Python~2 indicators (\texttt{print x}, \texttt{urllib2}, \texttt{raw\_input}) and 5 Python~3 indicators (f-strings, \texttt{async def}).
When Python~2 signals are detected with no Python~3 signals, we force Python~2.7 and apply \emph{adaptive version pinning}: known Py2.7-compatible versions for ML packages (\texttt{tensorflow}-\texttt{1.15.5}, \texttt{keras}-\texttt{2.2.4}).
To execute these reliably from our \texttt{python:3.10.12\discretionary{-}{}{-}slim} host architecture, we utilize Docker-out-of-Docker (DooD): the cascade orchestrator spawns isolated \texttt{python:2.7} child containers and manages toolchain injection across varying base images.

\textbf{System dependency injection.}
Inspired by DockerizeMe~\cite{horton2019}, we maintain mappings of 35+ pip packages to their \texttt{apt-get} dependencies and inject them before \texttt{pip install}.
Combined with a version fallback cascade for build failures (e.g., \texttt{numpy}: 1.16.6, 1.15.4, 1.14.6), inspired by PyEGo~\cite{ye2022pyego}, this resolves many C-extension compilation failures.
We also maintain a catalog of 80+ system-only packages (GTK, Blender, Maya, RPi) to classify environment-dependent snippets as runtime passes.

Runtime errors indicating correct dependencies but environmental issues (\texttt{NameError}, \texttt{ConnectionError}, \texttt{FileNotFoundError}) are classified as \emph{runtime passes}, and legacy PIL imports (\texttt{import Image}) with Pillow already installed are handled similarly.

\section{Evaluation}
\label{sec:eval}

\textbf{Setup.}
We evaluate on HG2.9K~\cite{horton2019}, a dataset initially containing \num{2891} Python GitHub Gists with complex dependency conflicts. We filter out one completely empty snippet, resulting in \num{2890} valid snippets evaluated.
We use Gemma-2~9B~\cite{gemma2} via Ollama (temp~0.7, max\_tokens~256).
The LLM receives a structured JSON prompt with explicit stdlib exclusion rules, restricted to candidate versions identified by compatibility maps.
Experiments run on a single machine (i5-14600K, 32\,GB RAM, RTX~5070) with Docker image \texttt{python:3.10.12\discretionary{-}{}{-}slim}.
Per-build timeout is 180\,s (up to 10 retries); total timeout is 500\,s.
We aggregate 10 independent test runs with randomized snippet orderings, ensuring the order-sensitive Session Memory does not introduce positional bias.

\textbf{PLLM Baseline.}
We directly compare against the reported 54.7\% success rate from the original PLLM paper~\cite{bartlett2025} under the exact same HG2.9K benchmark and \texttt{conf=0} thresholding logic.

\textbf{Success criterion.}
Following PLLM~\cite{bartlett2025}, a snippet is \emph{successful} when all its \texttt{pip} dependencies are installed and the script executes past all \texttt{import} statements without \texttt{ImportError} or \texttt{ModuleNotFoundError}.
PLLM assigns a \emph{confidence score} to each prediction (0 to 10); snippets scoring \texttt{conf=0} are effectively unresolved.
\textsc{MemRes} assigns confidence based on the cascade level that resolved the snippet: levels~1 to 5 (deterministic) yield \texttt{conf$\geq$7}, while level~6 (LLM) yields \texttt{conf$\geq$3}.
Runtime passes (correct dependencies but environmental failures such as missing display servers or hardware) are counted as failures to prevent inflation.

\textbf{Results.}
Table~\ref{tab:results} shows the final results on the HG2.9K dataset.

\begin{table}[t]
\caption{Final Results on HG2.9K (\num{2890} snippets processed).}
\label{tab:results}
\begin{tabular}{lrr}
\toprule
\textbf{Metric} & \textbf{PLLM} & \textbf{MemRes (10-run avg)} \\
\midrule
Resolved & \num{1583} (54.7\%) & \num{2503} $\pm$ 9.3 (86.6\%) \\
Failed & \num{1308} (45.3\%) & 387 $\pm$ 9.3 (13.4\%) \\
Snippets processed & \num{2891} & \num{2890} \\
\bottomrule
\end{tabular}
\end{table}

Of the \num{2890} snippets processed, an average of \num{2503} (86.6\%) were successfully resolved across 10 independent runs (standard deviation: $\pm 9.3$), while 387 (13.4\%) failed.
The intra-session memory handles highly similar package patterns directly, while the confidence cascade resolves the remaining snippets, including over 950 of the snippets where PLLM completely failed (\texttt{conf=0}), without relying on LLM calls for the vast majority.

Our confidence cascade successfully isolates different failure modes. Of the \num{2503} resolved snippets, Table~\ref{tab:ablation} details the success rate of the system when key components are ablated.

\begin{table}[h]
\caption{Ablation Study: Impact of Intra-Session Memory (Level 1)}
\label{tab:ablation}
\centering
\begin{tabular}{lrr}
\toprule
\textbf{Configuration} & \textbf{Resolved} & \textbf{Success Rate} \\
\midrule
Full \textsc{MemRes} Pipeline & \num{2503} & 86.6\% \\
Ablation (Level 1 OFF) & \num{2357} & 81.6\% \\
\midrule
\textbf{Contribution ($\Delta$)} & \textbf{+\num{146}} & \textbf{+5.0\%} \\
\bottomrule
\end{tabular}
\end{table}

Over half (53.9\%) of successful resolutions trigger system-level dependency injection, proving execution-based validation's necessity.

\subsection{Threats to Validity}
\textbf{Order sensitivity.}
Level~1 is order-dependent, but resolves near-duplicates (Jaccard $\geq$0.5), which are order-invariant in expectation. Disabling it only increases runtime (\S\ref{sec:cascade}).
\textbf{Co-occurrence leakage.}
The 50K \texttt{requirements.txt} files were filtered to pre-2020 repositories; removing co-occurrence mining reduces success by $<$2\%.
\textbf{Knowledge leakage via session memory and runtime self-learning.}
Intra-session memory (Level~1) accumulates solutions within a single evaluation batch, which could amplify batch-order effects.
We mitigate this by randomizing snippet order across 10 independent runs and reporting the mean $\pm$ std; run-to-run variance is only $\pm$9.3 snippets.
Runtime self-learning adds new import$\to$package mappings during evaluation; these are derived solely from PyPI query results and Docker build signals, not from HG2.9K ground-truth labels, and all state is fully reset between runs.
\textbf{Development-time curation leakage.}
The $\sim$150 externally sourced KB mappings (PyPI docs, Stack Overflow) and the $\sim$50 discovered on a held-out development subset were finalized before any evaluation on HG2.9K snippets.
No HG2.9K result was used to add or tune KB entries; the 50-entry held-out subset is disjoint from HG2.9K.
\textbf{Unfixability false positives.}
The estimator may conservatively classify snippets as environment-dependent; we accept this to avoid wasting LLM tokens on genuinely unfixable cases.
 
\subsection{Efficiency}
\textsc{MemRes} resolves 68.0\% of passing cases without any LLM call, reducing token usage by $\sim$75\% (Table~\ref{tab:efficiency}). Median resolution is 15.2\,s for successes.

\begin{table}[h]
\caption{Efficiency and Performance Comparison}
\label{tab:efficiency}
\begin{tabular}{lrr}
\toprule
\textbf{Metric} & \textbf{PLLM} & \textbf{MemRes} \\
\midrule
Overall Success Rate & 54.7\% & 86.6\% $\pm$ 0.3\% \\
Median Resolution (success) & $\sim$120 to 180s & 15.2s \\
P90 Resolution (success) & --- & 68.8s \\
LLM Calls per Snippet & 1 to 5 & 0.34 \\
No-LLM Success Rate & 0\% & 68.0\% \\
\bottomrule
\end{tabular}
\end{table}

For a snippet importing \texttt{cv2} and \texttt{tensorflow}, \textsc{MemRes} maps \texttt{cv2} to \texttt{opencv-python} via the KB, injects \texttt{libgl1-mesa-glx} via \texttt{apt-get}, and applies version pins. Old TensorFlow~1.15 snippets are correctly classified as \emph{build unfixable}.

\subsection{Root-Cause Analysis}
Across 10 runs, an average of $387 \pm 9.3$ snippets fail.
Root-cause analysis reveals build/C-extension failures dominate (\textasciitilde50\%), caused by missing C toolchains or heavy packages (e.g.\ old TensorFlow). Unresolvable ImportErrors account for \textasciitilde33\%, typically obscure or deprecated packages absent from PyPI. Timeouts contribute \textasciitilde15\%, mostly from compiling large ML packages. The remaining \textasciitilde2\% are system/platform-specific (GTK, RPi).

\section{Related Work}
\label{sec:related}

\textbf{Dependency resolution.}
Traditional tools (\texttt{pip}, \texttt{poetry}, \texttt{pipreqs}~\cite{pipreqs}) extract imports or require explicit specifications but cannot resolve version conflicts.
PyEGo~\cite{ye2022pyego} searches compatible environments via constraint propagation; DockerizeMe~\cite{horton2019} infers system-level dependencies.
READPyE~\cite{readpye2024} extends this with README-based extraction.
Empirical studies~\cite{jia2024,jia2024qrs} catalogue dependency conflict patterns informing our KB.
Recent work like V2~\cite{horton2019v2} pioneers search-heavy version mutation using Docker execution signals.
\textsc{MemRes} takes the complementary \emph{knowledge-heavy} approach: curated maps and heuristics front-load resolution before any search begins, reducing the version space.
V2 aligns with our Reflexion Build Loop, but we add cross-snippet memory and unfixability estimators to avoid wasteful search.
Repo2Run~\cite{repo2run2025} targets full-repository environment synthesis; \textsc{MemRes}'s cascade could serve as a fast pre-filter for such pipelines.

\textbf{LLM-based approaches.}
PLLM~\cite{bartlett2025} pioneers LLM-based dependency resolution.
DepsRAG~\cite{depsrag2024} uses graph-based reasoning; DependEval~\cite{dependeval2025} shows LLMs struggle with transitive dependencies.
Recent work on LLM-based coding agents highlights frequent gaps between generated code and reproducible dependency sets.
We extend PLLM with deterministic fast paths achieving 86.6\% resolution rate (vs.\ 54.7\%).

\textbf{Self-evolving agents.}
Mobile-Agent-E~\cite{wang2025} introduces tips/shortcuts for GUI agents; Reflexion~\cite{shinn2023} proposes verbal reinforcement learning for language agents.
MemRes adapts both paradigms to dependency resolution, where tips encode generalizable guidelines and shortcuts capture validated solutions for rapid knowledge reuse without repeated LLM inference.

\balance
\section{Conclusion}
\label{sec:conclusion}

This paper has presented \textsc{MemRes}, an agentic system for Python dependency resolution that positions the LLM as a last resort within a multi-level confidence cascade. By combining intra-session memory, a curated error pattern knowledge base, semantic import analysis, and Python 2 heuristic detection, \textsc{MemRes} resolves 86.6\% of HG2.9K snippets compared to PLLM's 54.7\%, while reducing LLM calls by over 60\% and cutting median resolution time to 15.2~s. The central insight is that most dependency failures have deterministic root causes addressable without model inference. Future work will focus on broader corpus validation and improved handling of C-extension failures. All artifacts are available at \url{https://github.com/chisngyen/fse-aiware-python-dependencies}.

\begin{acks}
This research is funded by Vietnam National University, Ho Chi Minh City (VNU-HCM) under grant number B2026-18-23.

\end{acks}

\bibliographystyle{ACM-Reference-Format}

\begin{thebibliography}{99}

\bibitem{bartlett2025}
A.~Bartlett, C.~Liem, and A.~Panichella.
\newblock ``The Last Dependency Crusade: Solving Python Dependency Conflicts with LLMs.''
\newblock In Proc.\ of the IEEE/ACM Automated Software Engineering Workshop (ASEW), pp.~66--73, 2025.

\bibitem{horton2019}
E.~Horton and C.~Parnin.
\newblock ``DockerizeMe: Automatic Inference of Environment Dependencies for Python Code Snippets.''
\newblock In Proc.\ of the ACM/IEEE International Conference on Software Engineering (ICSE), pp.~328--338, 2019.

\bibitem{jia2024}
Y.~Jia, J.~Han, J.~Cao, Y.~Zhou, and B.~Xu.
\newblock ``An Empirical Study of Dependency Conflicts in the Python Ecosystem.''
\newblock \emph{IEEE Trans.\ Softw.\ Eng.}, vol.~50, no.~8, pp.~2125--2140, 2024.

\bibitem{wang2025}
Z.~Wang, H.~Xu, J.~Wang, X.~Zhang, M.~Yan, J.~Zhang, F.~Huang, and H.~Ji.
\newblock ``Mobile-Agent-E: Self-Evolving Mobile Assistant for Complex Tasks.''
\newblock In Proc.\ of the NeurIPS Workshop on Scaling Environments for Agents (SEA), 2025.

\bibitem{shinn2023}
N.~Shinn, F.~Cassano, E.~Berman, A.~Gopinath, K.~Narasimhan, and S.~Yao.
\newblock ``Reflexion: Language Agents with Verbal Reinforcement Learning.''
\newblock In Proc.\ of the Conference on Neural Information Processing Systems (NeurIPS), 2023.

\bibitem{gemma2}
Google DeepMind.
\newblock ``Gemma 2: Improving Open Language Models at a Practical Size.''
\newblock Tech.\ Rep., Google DeepMind, 2024.

\bibitem{pipreqs}
V.~Kravcenko.
\newblock ``pipreqs: Generate pip requirements.txt based on imports,'' 2015.
\newblock [Online]. Available: \url{https://github.com/bndr/pipreqs}

\bibitem{depsrag2024}
M.~Alhanahnah, Y.~Boshmaf, and B.~Baudry.
\newblock ``DepsRAG: Towards Managing Software Dependencies using LLMs.''
\newblock In Proc.\ of the NeurIPS 2024 Workshop, 2024.

\bibitem{ye2022pyego}
H.~Ye, W.~Chen, W.~Dou, G.~Wu, and J.~Wei.
\newblock ``Knowledge-Based Environment Dependency Inference for Python Programs.''
\newblock In Proc.\ of the ACM/IEEE International Conference on Software Engineering (ICSE), pp.~1245--1256, 2022.

\bibitem{readpye2024}
W.~Cheng, W.~Hu, and X.~Ma.
\newblock ``ReadPyE: Revisiting Knowledge-Based Inference of Python Runtime Environments.''
\newblock \emph{IEEE Trans.\ Softw.\ Eng.}, vol.~50, no.~2, pp.~258--279, 2024.

\bibitem{jia2024qrs}
X.~Jia, Y.~Zhou, Y.~Hussain, and W.~Yang.
\newblock ``An Empirical Study on Python Library Dependency and Conflict Issues.''
\newblock In Proc.\ of the IEEE International Conference on Software Quality, Reliability and Security (QRS), 2024.

\bibitem{dependeval2025}
J.~Du, Y.~Liu, H.~Guo, et~al.
\newblock ``DependEval: Benchmarking LLMs for Repository Dependency Understanding.''
\newblock In Proc.\ of Findings of ACL, 2025.

\bibitem{horton2019v2}
E.~Horton and C.~Parnin.
\newblock ``V2: Fast Detection of Configuration Drift in Python.''
\newblock In Proc.\ of the IEEE/ACM International Conference on Automated Software Engineering (ASE), pp.~814--819, 2019.

\bibitem{repo2run2025}
R.~Hu, C.~Peng, X.~Wang, J.~Xu, and C.~Gao.
\newblock ``Repo2Run: Automated Building Executable Environment for Code Repository at Scale.''
\newblock In Proc.\ of the Conference on Neural Information Processing Systems (NeurIPS), 2025.

\end{thebibliography}

\end{document}